\documentclass{article}

\usepackage[preprint]{neurips_2024}

\usepackage[utf8]{inputenc} % allow utf-8 input
\usepackage[T1]{fontenc}    % use 8-bit T1 fonts
\usepackage{hyperref}       % hyperlinks
\usepackage{url}            % simple URL typesetting
\usepackage{booktabs}       % professional-quality tables
\usepackage{amsfonts}       % blackboard math symbols
\usepackage{nicefrac}       % compact symbols for 1/2, etc.
\usepackage{microtype}      % microtypography
\usepackage{xcolor}         % colors
\usepackage{graphicx} 
\usepackage{enumitem}

\title{SeeSay: An Assistive Device for the Visually Impaired Using  Retrieval Augmented Generation}

\author{%
  Melody Yu\thanks{The project information and code is available at https://github.com/cskitty/seesay  
    } \\
  Sage Hill School\\
  Newport Coast, CA 92657 \\
  \texttt{ocmelodyu@gmail.com} \\
}

\begin{document}

\maketitle

\begin{abstract}
In this paper, we present SeeSay, an assistive device designed for individuals with visual impairments. This system leverages large language models (LLMs) for speech recognition and visual querying. It effectively identifies, records, and responds to the user's environment by providing audio guidance using retrieval-augmented generation (RAG). Our experiments demonstrate the system's capability to recognize its surroundings and respond to queries with audio feedback in diverse settings. We hope that the SeeSay system will facilitate users' comprehension and recollection of their surroundings, thereby enhancing their environmental perception, improving navigational capabilities, and boosting overall independence.
\end{abstract}

\section{Introduction}

Visual impairment is a global health issue that affects over 253 million people, with 217 million experiencing moderate to severe vision loss, including 36 million who are blind [1]. Visual impairment severely hinders daily activities such as navigating surroundings, accessing information, and performing daily tasks. While canes help detect physical obstacles and navigate through environments, they cannot identify specific objects, read signs, or provide visual details that are farther away. Similarly, guide dogs require extensive training and care, face navigational limitations, and may not suit everyone due to health, legal, or personal reasons. There is an increasing need for more sophisticated solutions that offer accurate, real-time data about surroundings. 

Assistive technology that leverages deep learning models offers significant potential for addressing the challenges visually impaired individuals face. This technology provides real-time environmental information by utilizing speech recognition and image description techniques. Additionally, the capability to run large models on affordable, readily available devices, such as Raspberry Pis, presents a cost-effective approach to developing assistive devices for the visually impaired. This affordability makes these advanced technologies more accessible to a wider audience, including those with restricted financial resources.  

In this study, we introduce SeeSay, an assistive device for the visually impaired. It features low-cost components and leverages large language models to process images captured by a Bluetooth camera which can be installed on any glasses. The system uses a Raspberry Pi to deliver real-time audio feedback, enhancing the user's ability to navigate unfamiliar environments with greater confidence and independence. The performance of the SeeSay platform has been tested through various experiments, the details of which are discussed in subsequent sections of this paper.

\section{Related Work}

Assistive technology for the visually impaired is rapidly growing with diverse solutions available:

\textbf{Assistive Devices.} The studies in [2,3] describe systems employing sonar devices to detect obstacles within a short range and communicate this data to the user through real-time alerts, aiming to enhance user safety and facilitate more effective environmental navigation. The study detailed in [4] employed the Microsoft Kinect sensor alongside object recognition technology to offer real-time feedback to users about their surroundings. In [5], the authors described the development of a new augmented-reality application that can run on the Google Glass device. This application enhanced the real-world images by applying an edge enhancement filter, making the boundaries of objects clearer and easier to see. The authors state that this application can help people with low vision to see more clearly and thus improve their quality of life. The paper [6] described the design and implementation of a new navigation aid that integrates a camera, ultrasonic sensors, and a microcontroller to alert the user to obstacles and provide guidance on the best path to take. The paper [7] focused on the development and demonstration of a novel approach to control iOS's VoiceOver through a tactile button input system, making it easier for visually impaired individuals to use their mobile devices. In [8], the authors proposed a deep learning-based assistive system integrating an RGBD camera, earphones, and a smartphone interface to improve environmental perception for visually impaired individuals by providing walkable instructions and 3D spatial understanding through semantic mapping and touch interactions.  The paper [9] proposed an Open Scene Understanding (OpenSU) system that generates pixel-wise dense segmentation masks of entities involved in grounded situation recognition tasks, built on top of a pure transformer backbone with efficient segment modeling to improve scene understanding and facilitate independent mobility for visually impaired individuals.
 
 \textbf{Mobile Apps.} Microsoft's Seeing AI [10], is a free mobile app leveraging computer vision and machine learning, that narrates the surroundings for blind and visually impaired users, assisting with tasks such as reading text and identifying objects. Be My AI App [11] is a mobile application that can use visual querying models to provide accessibility guidance for blind or low-vision users with a success rate of 90\%.

Current visual querying devices for the visually impaired lack the ability to use past observations when providing new information, limiting their effectiveness in locating previously seen objects. This paper addresses this limitation by proposing a novel method that combines large language models and retrieval augmented generation to answer user queries using both current and past visual information.

\section{Methodology}

The SeeSay system consists of two primary components: a glasses attachment and a processing unit. First, the attachment comprises a compact 3D-printed enclosure designed to attach to the side of any pair of glasses, which houses an ESP32 board, a camera, and a battery. This assembly is engineered to capture photos and record audio, subsequently transmitting this data to the processing unit via Bluetooth. The second component, the processing unit, powered by a Raspberry Pi 5B, is responsible for analyzing both image and voice data and provides auditory feedback to the user.

To assist visually impaired users in navigating their environments, the SeeSay system persistently records their surroundings. It later utilizes Retrieval-Augmented Generation (RAG) to access and process this stored information, which enhancing its ability to provide more contextually relevant responses and support daily activities. SeeSay is open to a wide range of user inquiries:

\begin{itemize}
    \item \textbf{Scene Description:} Verbally describe the scene to the user and answer the user's questions, aiding in spatial awareness and environmental context. "Describe what you see."

    \item \textbf{Navigation Assistance:} Assist users in navigating unfamiliar settings. "Which direction should I take to find the restroom?"
    
    \item \textbf{Locate Items:} Help the user to locate items from their past locations, answering questions such as "Where did I leave my phone?" 
       
    \item \textbf{Optical Character Recognition (OCR):} Read and convert printed text into spoken words, allowing users to easily access written information in books, signs, and screens. "What is the first item on this menu?"
 
\end{itemize}

\begin{figure}
  \centering
  \includegraphics[width=0.9\linewidth]{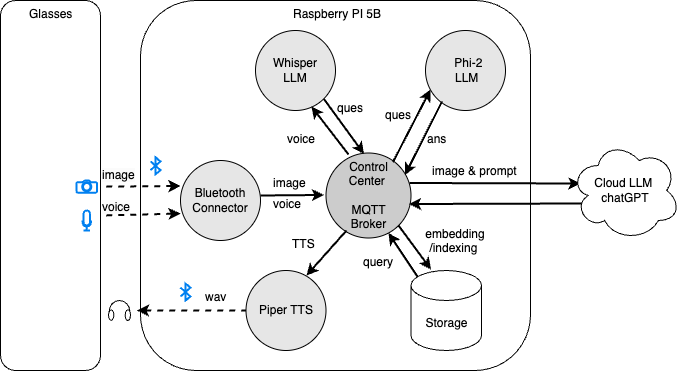} 
  \caption{Architecture of the SeeSay platform using LLM-based Retrieval-Augmented Generation}
  \label{fig:architecture}
\end{figure} 

The system utilizes a hybrid architecture to run large language models (LLMs). It runs the Whisper, Phi-2, and Piper TTS models locally on Raspberry Pi, handling tasks such as speech recognition, query answering, and speech synthesis, and delivering audio feedback to the user. For image description tasks, the system compensates for the Raspberry Pi's limited computational power by offloading the processing to a cloud-based LLM service such as ChatGPT. This involves transmitting the images and question prompts to the cloud for answering.

A control center orchestrates the system's control flow, while a mosquitto MQTT broker facilitates the publication and subscription of messages within the system. Upon receiving images and voice data from the glasses through the Bluetooth connection, the data is immediately published to the MQTT broker. Subsequently, the data is relayed to the relevant components for processing, establishing a continuous interaction loop that allows the user to pose sequential questions. The following is a detailed explanation of a user's question-and-answer flow within this system.

\begin{enumerate}[label=\textbf{\arabic*.}]

\item   The glasses capture a new image every 30 seconds and transmit it via Bluetooth to a Raspberry Pi for storage.
\item   The Raspberry Pi transmits the images to a cloud-based language model ChatGPT4 to obtain textual descriptions of the images. Image descriptions are transformed into vectors via embedding and stored locally with their corresponding images.
\item   The user issues a voice command and then the glasses transmit the audio recording to the Raspberry Pi, where the Whisper LLM processes the recording into a textual command.
\item   The control center executes multiple iterations of question-and-answer sessions with a local LLM (Dolphin 2.6 Phi 2) to process the query.
\begin{itemize}
        \item For simple questions, the local LLM  directly returns the answer.
        \item For inquiries related to the current environment, as detected by the local LLM model, the control center acquires the latest image description and combines it with the original question to re-query the local LLM for a response.
        \item For inquiries necessitating historical images, the local LLM creates a query string, and the control center retrieves the most relevant description (cosine similarity) from the database. This description is subsequently utilized as context to re-query the local LLM for an answer.
        \item If the response from the local LLM is inadequate, the user can request further assistance, and the control center will use ChatGPT to address the query, using the previously retrieved image as context.
        \item The user may issue a spoken command to request the inclusion of supplementary information to the latest image descriptions, thereby enhancing the contextual richness of the stored image description, "Remember this person as Mary."
    \end{itemize}
\item  The answer derived from the previous step is converted to speech using Piper TTS and relayed through Bluetooth-connected speakers or headphones for the user to hear.

\end{enumerate}

\section{Experimental Results}

\begin{table}[htbp]
\centering
\caption{Performance Metrics for Visual Assistive Device Test Cases}
\begin{tabular}{@{}lccc@{}}
\toprule
Test Case & Accuracy (\%) & Usability (1-5) & Response Time (s) \\ \midrule
Simple Question & 86 & 4.5 & 9.2 \\
Scene Description & 86 & 4.5 & 10.5 \\
Indoor Navigation & 70 & 3.5 & 18.5 \\
Street Navigation & 70 & 3.0 & 16.5 \\
Image Inquiries & 80 & 3.5 & 17.5 \\
Recognize Person & 85 & 4.0 & 16.5 \\
Item Locator & 80 & 4.0 & 18.8 \\
Printed OCR & 85 & 4.5 & 10.3 \\
Handwriting OCR & 75 & 4.0 & 11.5 \\
\bottomrule
\end{tabular}
\label{tab1e:result2}
\end{table}
 
To evaluate the SeeSay system, we tested the prototype design across various test scenarios, with each focusing on essential functionalities for the visually impaired. The device exhibited commendable performance in simpler tasks such as answering direct questions and describing scenes, achieving high accuracy and usability scores of 86\% and 4.5, respectively, with response times of 9.2 and 10.5 seconds. These metrics suggest that the device is effective in providing information retrieval and enhancing environmental awareness, essential for real-time decision-making.

However, more complex navigational tasks like indoor and street navigation presented some challenges, evidenced by lower accuracy of 70\%, and relatively long response times of 18.5 seconds. While the device performed reasonably well in recognizing individuals and processing image inquiries, the latter's longer response time (17.5 seconds) reflected multiple rounds of question and answer with the local LLM. Tasks such as item location and optical character recognition for printed text also demonstrated higher usability and satisfactory response times, highlighting the device's utility in aiding users to locate objects and read printed material efficiently. Handwriting OCR also performed adequately, though improvements could enhance its correctness and responsiveness. The experimental results reveal that the SeeSay system effectively supports visually impaired users in simple tasks and scene descriptions, but also suggest that improvements in navigation tasks and handwriting OCR could increase overall usability.

\section {Conclusion}

The prototype evaluations of the SeeSay system reveal its capability to deliver usable and efficient speech outputs. Nonetheless, the response times are delayed and fluctuate based on the intricacy of the question and answer interactions.  These outcomes highlight the system's capability to support various assistive tasks, demonstrating its potential as a viable tool for visually impaired users.

In this paper, we propose continuous capturing and indexing of users' visual observations to enrich the contextual data used by LLMs, thus extending the functionalities of visually assistive devices. This mechanism serves as a visual memory aid for the visually impaired, thus enhancing the system's capacity to respond to user queries according to previously observed information. Through the SeeSay prototype, we have shown that integrating Retrieval Augmented Generation with Large Language Models can create a user-friendly, natural language-based assistance device. This integration improves the accuracy and relevance of interactions for users of visual assistance devices who rely on such technology for visual information.

Given the constrained processing capabilities and memory of the Raspberry Pi, tasks related to image description have to be delegated to cloud-based LLMs, thus elevating the operational costs. Future enhancements to the SeeSay system will involve integrating a more powerful computing platform capable of locally executing vision LLM models. This upgrade aims to reduce dependence on cloud-based LLM services and boost user privacy by enabling on-device LLM functionality.

\section*{References}

[1] Ackland, P., Resnikoff, S.\ \& Bourne R.\ (2017) World blindness and visual impairment: despite many successes, the problem is growing. {\it Community Eye Health }. 2017;30(100):71-73.  

[2] Dunai, L., Fajarnes, G. P.,  Praderas, V. S., Garcia, B. D. and Lengua, I. L. (2010) "Real-time assistance prototype — A new navigation aid for blind people," IECON 2010 - 36th Annual Conference on IEEE Industrial Electronics Society, Glendale, AZ, USA, 2010, pp. 1173-1178 

[3] Kamaludin, M.H., Mahmood, N.H., Ahmad, A.H. and Omar C.(2015), ``Sonar assistive device for visually impaired people'' in Jurnal Teknologi 73.6 . 

[4] Takizawa, H, Yamaguchi,S., Aoyagi, M., Ezaki, N and Mizuno, S (2012) ``Kinect cane: An assistive system for the visually impaired based on the concept of object recognition aid'' in  Personal and Ubiquitous Computing 19: pp. 955-965.

[5] Hwang, A. D., and Peli, E. (2014) "An augmented-reality edge enhancement application for Google Glass." Optometry and vision science 91, no. 8: 1021-1030.

[6] Mustapha, B., Zayegh, A. and Begg, R. K. (2013) "Wireless obstacle detection system for the elderly and visually impaired people." In 2013 IEEE International Conference on Smart Instrumentation, Measurement and Applications (ICSIMA), pp. 1-5. 

[7] Batterman, J. M., Martin, V.F., Yeung, D., and Walker, B.N  (2018) "Connected cane: tactile button input for controlling gestures of iOS voiceover embedded in a white cane." Assistive Technology 30, no. 2: pp. 91-99.

[8] Lin, Y., Wang, K., Yi,  W. and Lian, S. (2019) “Deep learning based wearable
assistive system for visually impaired people,” in Proceedings of the
IEEE/CVF international conference on computer vision workshops

[9] Liu, R., Zhang, J. , Peng, K. , Zheng, J. , Cao, K. , Chen, Y., Yang,  K. and Stiefelhagen, R. (2023) “Open scene understanding: Grounded situation
recognition meets segment anything for helping people with visual
impairments,” in Proceedings of the IEEE/CVF International Conference
on Computer Vision, pp. 1857–1867 

[10] Microsoft, Seeing AI: New Technology Research to Support the Blind and Visually Impaired Community, https://blogs.microsoft.com/accessibility/seeing-ai/

[11] Be My AI https://www.bemyeyes.com/blog/introducing-be-my-ai

\end{document}